



\input lanlmac
\input amssym

\newcount\figno
\figno=0
\def\fig#1#2#3{
\par\begingroup\parindent=0pt\leftskip=1cm\rightskip=1cm\parindent=0pt
\baselineskip=11pt
\global\advance\figno by 1
\midinsert
\epsfxsize=#3
\centerline{\epsfbox{#2}}
\vskip 12pt
\centerline{{\bf Fig. \the\figno:~~} #1}\par
\endinsert\endgroup\par
}
\def\figlabel#1{\xdef#1{\the\figno}}


\def\th{\theta}

\def\ep{\epsilon}
\def\vep{\varepsilon}

\def\hf{{1\over 2}}
\def\qu{{1\over 4}}

\def\o{\over}
\def\til#1{\widetilde{#1}}
\def\si{\sigma}

\def\b#1{\overline{#1}}
\def\del{\partial}

\def\lap{\Delta}
\def\bra{\langle}
\def\ket{\rangle}
\def\lf{\left}
\def\ri{\right}
\def\riya{\rightarrow}

\def\Riya{\Rightarrow}

\def\la{\lambda}
\def\La{\Lambda}

\def\Ga{\Gamma}

\def\dag{\dagger}
\def\rt#1{\sqrt{#1}}

\def\sitarel#1#2{\mathrel{\mathop{\kern0pt #1}\limits_{#2}}}

\lref\SeibergEI{
N.~Seiberg and D.~Shih,
``Flux vacua and branes of the minimal superstring,''
JHEP {\bf 0501}, 055 (2005)
[arXiv:hep-th/0412315].
}
\lref\KutasovFG{
  D.~Kutasov, K.~Okuyama, J.~w.~Park, N.~Seiberg and D.~Shih,
  ``Annulus amplitudes and ZZ branes in minimal string theory,''
  JHEP {\bf 0408}, 026 (2004)
  [arXiv:hep-th/0406030].
}
\lref\MaldacenaSN{
  J.~Maldacena, G.~W.~Moore, N.~Seiberg and D.~Shih,
  ``Exact vs. semiclassical target space of the minimal string,''
  JHEP {\bf 0410}, 020 (2004)
  [arXiv:hep-th/0408039].
}
\lref\MorozovHH{
  A.~Morozov,
  ``Integrability and matrix models,''
  Phys.\ Usp.\  {\bf 37}, 1 (1994)
  [arXiv:hep-th/9303139].
}
\lref\SeibergNM{
  N.~Seiberg and D.~Shih,
  ``Branes, rings and matrix models in minimal (super)string theory,''
  JHEP {\bf 0402}, 021 (2004)
  [arXiv:hep-th/0312170].
}
\lref\FukudaBV{
  T.~Fukuda and K.~Hosomichi,
  ``Super Liouville theory with boundary,''
  Nucl.\ Phys.\ B {\bf 635}, 215 (2002)
  [arXiv:hep-th/0202032].
}
\lref\AhnEV{
  C.~Ahn, C.~Rim and M.~Stanishkov,
  ``Exact one-point function of N = 1 super-Liouville theory with boundary,''
  Nucl.\ Phys.\ B {\bf 636}, 497 (2002)
  [arXiv:hep-th/0202043].
}
\lref\MartinecKA{
  E.~J.~Martinec,
  ``The annular report on non-critical string theory,''
  arXiv:hep-th/0305148.
}
\lref\GaiottoYB{
  D.~Gaiotto and L.~Rastelli,
  ``A paradigm of open/closed duality: Liouville D-branes and the  Kontsevich
  model,''
  arXiv:hep-th/0312196.
}
\lref\HashimotoBF{
  A.~Hashimoto, M.~x.~Huang, A.~Klemm and D.~Shih,
  ``Open / closed string duality for topological gravity with matter,''
  arXiv:hep-th/0501141.
}
\lref\KlebanovWG{
  I.~R.~Klebanov, J.~Maldacena and N.~Seiberg,
  ``Unitary and complex matrix models as 1-d type 0 strings,''
  Commun.\ Math.\ Phys.\  {\bf 252}, 275 (2004)
  [arXiv:hep-th/0309168].
}
\lref\JohnsonHY{
  C.~V.~Johnson,
  ``Non-perturbative string equations for type 0A,''
  JHEP {\bf 0403}, 041 (2004)
  [arXiv:hep-th/0311129].
}
\lref\JohnsonUT{
  C.~V.~Johnson,
  ``Tachyon condensation, open-closed duality, resolvents, and minimal bosonic
  and type 0 strings,''
  arXiv:hep-th/0408049.
}
\lref\GaiottoNZ{
  D.~Gaiotto, L.~Rastelli and T.~Takayanagi,
  ``Minimal superstrings and loop gas models,''
  arXiv:hep-th/0410121.
}
\lref\KapustinPK{
  A.~Kapustin,
  ``A remark on worldsheet fermions and double-scaled matrix models,''
  arXiv:hep-th/0410268.
}
\lref\KawaiPJ{
  H.~Kawai, T.~Kuroki and Y.~Matsuo,
  ``Universality of nonperturbative effect in type 0 string theory,''
  arXiv:hep-th/0412004.
}
\lref\CarlisleMK{
  J.~E.~Carlisle, C.~V.~Johnson and J.~S.~Pennington,
  ``Baecklund transformations, D-branes, and fluxes in minimal type 0
  strings,''
  arXiv:hep-th/0501006.
}

\Title{             
                                             \vbox{
                                             \hbox{hep-th/0503082}}}
{\vbox{
\centerline{Annulus Amplitudes in the Minimal Superstring}
}}

\vskip .2in

\centerline{Kazumi Okuyama}

\vskip .2in

\centerline{Department of Physics and Astronomy, 
University of British Columbia} 
\centerline{Vancouver, BC, V6T 1Z1, Canada}
\centerline{\tt kazumi@physics.ubc.ca}
\vskip 3cm
\noindent

%
We study the annulus amplitudes in the $(2,4)$ minimal superstring
theory using the continuum worldsheet approach.
Our results reproduce the semiclassical 
behavior of the wavefunctions of FZZT-branes recently studied  in 
hep-th/0412315 using the dual matrix model. 
We also study the multi-point functions
of neutral FZZT-branes
and find the agreement
between their semiclassical limit and the worldsheet 
annulus calculation.

\Date{March, 2005}

\vfill
\vfill

\newsec{Introduction}
Minimal superstring theories are interesting arena to study
various aspects of string theory \refs{\KlebanovWG\JohnsonHY\SeibergNM\JohnsonUT\GaiottoNZ\KapustinPK\KawaiPJ{--}\CarlisleMK}. 
The existence of dual matrix models allows us to study 
nonperturbative phenomena in a very controlled setup.
In particular, we can expect to understand the physics of D-branes
in a quantum regime and the open-closed duality in a very precise way.

In \MaldacenaSN\ it is realized that the FZZT-brane is a useful probe
of the target space in minimal string theories. In the matrix model 
description, the FZZT-brane corresponds to the determinant operator
$\det(x-M)$ and one can define the wavefunction $\psi(x)$ of FZZT-brane
as a double scaling limit of the expectation value $\bra\det(x-M)\ket$.
Due to the Stokes' phenomenon, the semiclassical moduli space of 
FZZT-brane, which is represented by a Riemann surface, is replaced
by the entire complex $x$-plane in the full theory,
and correspondingly 
$\psi(x)$ is an entire function of $x$. 
In particular, at a topological point 
the FZZT wavefunction is given by a (generalized)
Airy function \refs{\MaldacenaSN,\HashimotoBF,\GaiottoYB}
which is indeed entire.
This implies that a semiclassical picture of target space measured by
the FZZT-brane is completely different from the picture in the
full nonperturbative theory.

One would expect that the physics of
FZZT-branes and their wavefunctions in minimal superstring theories
is much richer than that of their bosonic cousins, 
given the fact that there is an interesting interplay 
between RR fluxes and D-branes in minimal superstring theories. 
In a recent paper \SeibergEI, the simplest minimal superstring, {\it i.e.},
the $(p,q)=(2,4)$ theory has been studied. 
This theory contains two types of FZZT-branes, the branes which are
either neutral or charged under the RR potential.
The neutral brane is easy to describe in the 0A language
(complex matrix model), while the charged brane is naturally
described in the 0B language (hermitian matrix model).
It is argued \KlebanovWG\ that 0A and 0B theories are actually
equivalent for the $(p,q)=(2,4)$ case.

In this paper, we will present a one more nontrivial check for 
the duality between the matrix model 
and the $(2,4)$ minimal superstring theory.
We will study the annulus amplitudes in the $(2,4)$ theory using
the worldsheet approach
and find complete agreement with the semiclassical behavior of the
wavefunctions of FZZT-branes obtained in \SeibergEI.
Although the wavefunctions are analytic in the bulk cosmological constant
$\mu$, the semiclassical limit is dependent on the sign of $\mu$,
so we need to study the two cases 
$\mu>0$ and $\mu<0$ separately.

The semiclassical expansion of FZZT wavefunction
involves only the annulus between the FZZT-brane and itself.
On the other hand, our worldsheet calculation gives more general
amplitudes between two FZZT-branes with different boundary cosmological
constants $\mu_B\not=\mu_B'$.
These amplitudes naturally appear in the semiclassical limit
of multi-point functions of FZZT-branes.
We will study the multi-brane correlators of neutral FZZT-branes
using the complex matrix model
and again find an impressive agreement with the worldsheet calculation.

This paper is organized as follows.
In section 2, we study the annulus amplitudes in the $\mu>0$ phase
and compare them with  the semiclassical behavior of the wavefunctions
of FZZT-branes.
In section 3, we repeat the analysis for the $\mu<0$ phase.
In section 4, we consider the multi-point functions of neutral FZZT-branes
and show that the semiclassical behavior of those correlators
is reproduced from the worldsheet annulus computation.
Appendix A is devoted to the detailed calculation of
the amplitudes.

\newsec{One-Cut Phase $(\mu>0)$}
Let us first summarize the semiclassical behaviors of the wavefunctions
of FZZT branes obtained in \SeibergEI. In this section, we will consider the
case $\mu>0$, or the one-cut phase in the type 0B picture.
In general, we expect that in the semiclassical regime ($\mu\riya+\infty$,
 $x\riya \pm i\infty$ with $x/\rt{\mu}$ fixed)
the wavefunction 
of FZZT-brane behaves as
\eqn\generalpsi{
\psi(x,q)\sim 
\exp\lf[D(x)+\hf Z(x,x)+|q|\til{Z}(x,\ep)+\cdots\ri]~,
}
where $x$ is related to the boundary cosmological constant $\mu_B$
by $x=i\mu_B$.
The first term $D(x)$ in \generalpsi\ denotes the disk amplitude.
The explicit form of disk for the charged FZZT-brane $D_{\pm}(x)$
and for the neutral FZZT-brane $D_0(x)$ reads
\eqn\diskD{
D_\pm(x)=\lf\{
\eqalign{&-i\lf({4\o3}x^3+\mu x\ri)\cr 
&-i{4\o3}\lf(x^2+{\mu\o2}\ri)^{{3\o2}}}\ri.~,
\quad
D_0(x)=\lf\{
\eqalign{&-{4\o3}\rt{2}\lf({\mu\o2}-x^2\ri)^{3\o2}\cr 
&-i\rt{2}\lf({4\o3}x^3-\mu x\ri)}\ri.
\quad\eqalign{
&\mu>0\cr &\mu<0}~.
}
The second term $Z(x,x)$ in \generalpsi\
is the annulus between the FZZT-brane and itself.
The factor of $1/2$ in front of $Z(x,x)$ comes from the fact that the
open strings are ending on the same brane \MaldacenaSN .
The third term $|q|\til{Z}(x,\ep)$ comes from the interaction between
the FZZT-brane and the $(1,1)$ ZZ-branes in the background.
At the leading order, this interaction is represented
by the annulus between the FZZT-brane and a single 
ZZ-brane with charge $\ep={\rm sign}(q)$, multiplied by the number $|q|$
of ZZ-branes.

It is shown in \SeibergEI\ that the semiclassical wavefunction of the
neutral FZZT-brane is given by
\eqn\psizeroone{
\psi_0(x,q)\sim (-ix)^{-|q|}\exp\lf[D_0(x)-\qu\log(\mu-2x^2)+{|q|\o2}
\log\lf({\rt{\mu-2x^2}-\rt{\mu}\o\rt{\mu-2x^2}+\rt{\mu}}\ri)+\cdots\ri]
}
and the wavefunction of the charged FZZT-brane behaves as
\eqn\psipmone{\eqalign{
\psi_+(x,q)&\sim\exp\lf[D_+(x)+q\log\big\{-i2\rt{2}(x-x_{ZZ})\big\}+\cdots\ri]\quad
{\rm Im}x>0\cr
\psi_-(x,q)&\sim\exp\lf[D_-(x)-q\log\big\{+i2\rt{2}(x-x_{ZZ})\big\}+\cdots\ri]\quad
{\rm Im}x<0~,
}}
where $x_{ZZ}=\mp i{\rt{\mu}\o2}$ is the value of $x$ associated with
the $(1,1)$ ZZ-brane, which corresponds to the extrema of the effective
potential in the dual matrix model picture.
For the neutral brane wavefunction \psizeroone, 
we have factored out $x^{-|q|}$ as an overall factor.
As discussed in \SeibergEI,
this
is necessary in the $\mu<0$ phase when comparing $\psi_0$ to the worldsheet
calculation.
Here we followed the same prescription also for the $\mu>0$
phase in order to identify the $q$-dependent term
$|q|\til{Z}(x,\ep)$ correctly.
In other words, the rescaled wavefunction 
$\til{\psi}_0(x)=(-ix)^{|q|}\psi_0(x)$
is the one which shows the expected semiclassical
behavior \generalpsi.

In the rest of this section, 
we will reproduce these expressions from the
continuum worldsheet calculation.

\subsec{Worldsheet Computation of the Annulus between FZZT Branes}
In the worldsheet approach, D-branes are described by
using boundary states.
The boundary states of the charged and neutral FZZT-branes are
given by\foot{In this paper, we use the normalization 
of $P$ and $\si$ in \SeibergNM, while the convention of
$\mu$ is that of \SeibergEI.} \refs{\SeibergNM,\SeibergEI}
\eqn\Bstateone{\eqalign{
&|\si,\xi,\eta=-1\ket=\int_0^\infty dP\lf({\mu\o2}\ri)^{iP\o b}
\Big(\cos(\pi P\si)A_{NS}(P)|P,\eta=-1\ket_{NS} \cr
&\hskip45mm+\xi\cos(\pi P\si)
A_R(P)|P,\eta=-1\ket_R\Big) \cr
&|\si,0,\eta=+1\ket=\rt{2}\int_0^\infty dP\lf({\mu\o2}\ri)^{iP\o b}
\cos(\pi P\si)A_{NS}(P)
|P,\eta=+1\ket_{NS}
}}
where $b={1\o\rt{2}}$ and $A_{NS}(P),A_R(P)$ are the wavefunctions of
${\cal N}=1$ super-Liouville theory \refs{\FukudaBV,\AhnEV}
\eqn\ANSR{
A_{NS}(P)={\Ga(1-iPb)\Ga(1-iPb^{-1})\o-\rt{2}\pi i P},\quad
A_{R}(P)={1\o\pi b^2}\Ga\lf(\hf-iPb\ri)\Ga\lf(\hf-iPb^{-1}\ri)~.
}
In \Bstateone, $\xi=\pm1$ represents the charge of the brane
and $\eta=\pm1$ specifies the boundary condition
of the supercharge $Q=i\eta\b{Q}$.

Now let us consider the annulus amplitudes between 
FZZT-branes
\eqn\annBB{\eqalign{
Z(\si,0|\si',0)&=\int_0^\infty dt_{c}\,
\bra \si,0,\eta=+1|q^{\hf(L_0+\til{L}_0)}|
\si',0,\eta=+1\ket \cr
Z(\si,\xi|\si',\xi)&=
\int_0^\infty dt_{c}\,\bra \si,\xi,\eta=-1|q^{\hf(L_0+\til{L}_0)}|
\si',\xi,\eta=-1\ket \cr
Z(\si,0|\si',\xi)&=-
\int_0^\infty dt_{c}\,\bra \si,0,\eta=+1|q^{\hf(L_0+\til{L}_0)}|
\si',\xi,\eta=-1\ket
}}
where $q=e^{-2\pi t_c}$.
In the last equation in \annBB, we have 
put an extra minus sign for the amplitude 
$Z(\si,0|\si',\xi)$,
since the modes running in the open string channel are
in the R-sector and hence fermionic.
From the calculations in appendix A,
the annulus amplitudes between various FZZT-branes in the $\mu>0$ phase
are found to be
\eqn\FZZTFZZT{\eqalign{
Z(\si,0|\si',0)&=-\log\Big(2\rt{\mu}\cosh\th+2\rt{\mu}\cosh\th'\Big) \cr
Z(\si,\xi|\si',\xi')&={-1+\xi\xi'\o2}
\log\Big(2\rt{\mu}\cosh\th+2\rt{\mu}\cosh\th'\Big) \cr
Z(\si,0|\si',\xi)&=
\hf\log\lf(2\rt{\mu}\cosh(\th+\th')+\rt{2\mu}\ri)
+\hf\log\lf(2\rt{\mu}\cosh(\th-\th')+\rt{2\mu}\ri)
}}
where we defined $\th$ by 
\eqn\thetavssigma{
\th={\pi b\si\o2}={\pi\si\o\rt{8}}~.
}

\subsec{$Z(x,x)$}
Now we can compare the worldsheet result \FZZTFZZT\ with the
matrix model result \psizeroone, \psipmone.
Let us consider the term $Z(x,x)$ in the semiclassical wavefunction.
For the charged branes,
the worldsheet result \FZZTFZZT\
shows that the annulus between FZZT-branes carrying the
same charge vanishes 
\eqn\Zxixivanish{
Z(\si,\xi|\si',\xi)=0~.
}
This agrees with the absence of the $q$-independent term 
in the wavefunctions \psipmone.

For the neutral brane, recalling that
$\th$ and $x$ are related by $x=i\rt{\mu\o2}\sinh\th$,
we see that the
worldsheet result \FZZTFZZT\ and the matrix model result
\psizeroone\ agree with each other up to an irrelevant additive constant
\eqn\Zxxzeroagree{
Z(\si,0|\si,0)=-\log(4\rt{\mu}\cosh\th)=-\hf\log(\mu-2x^2)-\log4~.
} 

\subsec{$\til{Z}(x,\ep)$}
Next consider the term $\til{Z}(x,q)$.
The annulus between a FZZT-brane and a ZZ-brane can be obtained from the 
FZZT-FZZT annulus by writing the ZZ boundary state
as a linear combination of the FZZT boundary states 
\refs{\SeibergNM,\MartinecKA}
\eqn\ZZvsFZZT{
|(1,1),\xi\ket=|\si_{1,1},\xi\ket-|\si_{1,-1},-\xi\ket
}
where $\si_{m,n}=i(mb^{-1}+nb)$. 
From this relation \ZZvsFZZT\ and the amplitude $Z(\si,0|\si',\xi)$
in \FZZTFZZT,
the annulus between a neutral FZZT-brane and a $(1,1)$ ZZ-brane 
is obtained as
\eqn\FZZTZZneu{
Z(\si,0|(1,1),\xi)=Z(\si,0|\si_{1,1},\xi)-Z(\si,0|\si_{1,-1},-\xi)
=\hf\log\lf({\cosh\th-1\o\cosh\th+1}\ri)~.
}
This is independent of the charge $\xi$ of the ZZ-brane
since only the NSNS exchange in the closed string channel contributes
to this amplitude.
From the relation $x=i\rt{\mu\o2}\sinh\th$, one can see that
\FZZTZZneu\ agrees with the $|q|$ dependent term inside
the exponential of \psizeroone.
Note that it is necessary to factor out $x^{-|q|}$
for this agreement to work.

Similarly, the annulus between a charged FZZT-brane and a $(1,1)$
ZZ-brane is given by
\eqn\FZZTZZch{\eqalign{
Z(\si,\xi|(1,1),\xi')&=Z(\si,\xi|\si_{1,1},\xi')-Z(\si,\xi|\si_{1,-1},-\xi') \cr
&={\xi\xi'-1\o2}\log\lf[2\rt{\mu}\Big(\cosh\th-{1\o\rt{2}}\Big)\ri]
+{\xi\xi'+1\o2}\log\lf[2\rt{\mu}\Big(\cosh\th+{1\o\rt{2}}\Big)\ri]. 
}}
Since $\xi\xi'=\pm1$, \FZZTZZch\ can be written more compactly as
\eqn\FZZTZZchcompact{
Z(\si,\xi|(1,1),\xi')=\xi\xi'\log\lf[2\rt{\mu}\Big(\cosh\th+{\xi\xi'\o\rt{2}}\Big)\ri]~.
}
Using the relation $x=i\xi\rt{\mu\o2}\cosh\th$ and
$x_{ZZ}=-i\xi'{\rt{\mu}\o2}$, \FZZTZZchcompact\ is further rewritten as
\eqn\ZvsY{
|q|Z(\si,\xi|(1,1),\xi')=2q_bq\log\Big[-i2\rt{2}\xi(x-x_{ZZ})\Big]
}
where we identified 
$\xi=2q_b={\rm sign}({\rm Im}x)$ and $\xi'={\rm sign}(q)$. 
This reproduces the $q$-dependent term
in the charged FZZT-brane wavefunctions \psipmone.
Although the final result is proportional to the charge of ZZ-brane, 
this amplitude receives contributions both from the RR sector 
and the NSNS sector.
However, there is a qualitative difference between the RR 
and the NSNS contributions
for this amplitude.
In the NSNS sector the pole of
$|A_{NS}(P)|^2$ is canceled by a factor coming from 
the ZZ boundary state
\eqn\nopoleinNS{
(\cos\pi P\si_{1,1}-\cos\pi P\si_{1,-1})|A_{NS}(P)|^2=1~.
}
Therefore, there is no pole at $P=0$ in the integral representation
of this NSNS contribution and the result is finite:
\eqn\NSexchZZ{
Z(\si,\xi|(1,1),\xi')_{NS}=\hf\int_{-\infty}^\infty{dP\o P}{\sinh{\pi P\o\rt{2}}
\o\cosh\rt{2}\pi P}\cos(\pi P\si)~.
}
On the other hand, the corresponding relation for the RR sector
\eqn\RRcancelA{
(\cos\pi P\si_{1,1}+\cos\pi P\si_{1,-1})|A_{R}(P)|^2=1
}
does not remove the pole at $P=0$:
\eqn\RRexchZZ{
Z(\si,\xi|(1,1),\xi')_{R}=-{\xi\xi'\o2}\int_{-\infty}^\infty{dP\o P}{\cosh{\pi P\o\rt{2}}
\o\sinh\rt{2}\pi P}\cos(\pi P\si)~.
}
Therefore, the RR contribution contains a divergence proportional
to the volume of Liouville direction.

Some comments on the statistics of open strings between
FZZT and ZZ branes are in order.
In the amplitude $Z(\si,0|(1,1),\xi)$, the open string running
along the loop is in the R sector. 
This implies that the open string between the
$(1,1)$ ZZ-brane and the neutral 
FZZT-brane is fermionic. This is consistent with the fact that
the neutral brane is represented by a determinant in the 
complex matrix model \SeibergEI.
On the other hand, $Z(\si,\xi|(1,1),\xi')$ is in the NS sector
in the open string channel
and hence the open string between the $(1,1)$ ZZ-brane and 
the charged FZZT-brane is bosonic.

This is in contrast with the situation in the bosonic minimal string.
In the bosonic theory, the FZZT-brane corresponds
to the determinant in the matrix model and consequently
the open strings 
between FZZT and ZZ branes are  fermionic, despite 
the fact that the theory
itself is a bosonic string theory.
In the worldsheet approach, 
this fermionic nature of open string can be taken care of
by adding an extra minus sign relative to the relation between
the FZZT boundary state and the ZZ boundary state 
in the Liouville theory alone \KutasovFG
\eqn\ZZinbosonic{
|m,n\ket=-|\si_{m,n}\ket+|\si_{m,-n}\ket~.
}
However, in the case of minimal superstring, 
this modification 
does not lead to a consistent picture
because the open string between ZZ and FZZT
can be either bosonic or fermionic depending on the
charge of FZZT-branes in question, as we saw above.
Instead, we keep the relation \ZZvsFZZT\ intact 
and add an extra minus sign to the annulus $Z(\si,0|\si',\xi)$ 
in order to represent the fermionic nature of the modes running along
the open string one-loop.
Our choice \ZZvsFZZT\ of the relative sign between ZZ and FZZT 
boundary states is justified by the fact that the disk amplitude
of the ZZ-brane is negative with this choice of relative sign
\foot{We would like to thank
N. Seiberg and D. Shih for this argument.} 
\eqn\ZZdisk{
D_{\pm}^{\rm ZZ}=
D_{\pm}(\si_{1,1})-D_{\mp}(\si_{1,-1})=-{2\o3}\mu^{3\o2} <0~.
}
This sign guarantees that the nonperturbative effect associated with
an instanton-anti-instanton pair, 
which goes like $e^{D_{+}^{\rm ZZ}+D_{-}^{\rm ZZ}}
\sim e^{-{4\o3}\mu^{3\o2}}$, is exponentially small in the weak coupling regime 
$g_s\sim\mu^{-3/2}\ll1$
\KutasovFG.

\newsec{Two-Cut Phase $(\mu<0)$}
The computation in the $\mu<0$ phase is almost similar to the $\mu>0$ case.
The FZZT boundary states in this phase are given by 
\refs{\SeibergNM,\SeibergEI}
\eqn\Bstatetwo{\eqalign{
&|\si,\xi,\eta=-1\ket^{naive}
=\int_0^\infty dP\lf(-{\mu\o2}\ri)^{iP\o b}
\Big(\cos(\pi P\si)A_{NS}(P)|P,\eta=-1\ket_{NS} \cr
&\hskip45mm-i\xi\sin(\pi P\si)
A_R(P)|P,\eta=-1\ket_R\Big) \cr
&|\si,0,\eta=+1\ket=\rt{2}\int_0^\infty dP\lf(-{\mu\o2}\ri)^{iP\o b}\cos(\pi P\si)A_{NS}(P)
|P,\eta=+1\ket_{NS}~.
}}
It was pointed out in \SeibergEI\ 
that for the charged FZZT-brane the coupling to the zero-momentum RR potential
should be added to the naive boundary state
\eqn\FZZTcorrect{
|\si,\xi,\eta=-1\ket=|\si,\xi,\eta=-1\ket^{naive}+{\xi\o2}V_R|0\ket~.
}

The annulus amplitudes $Z(\si,0|\si',0)$ and
$Z(\si,0|\si',\xi)$ involving neutral branes
are obtained from the $\mu>0$ result
by simply sending $\mu\riya-\mu$, since the RR sector does not contribute
to those amplitudes:
\eqn\Zneuttwo{\eqalign{
Z(\si,0|\si',0)&=-\log\Big(2\rt{-\mu}\cosh\th+2\rt{-\mu}\cosh\th'\Big) \cr
Z(\si,0|\si',\xi)&=
\hf\log\lf(2\rt{-\mu}\cosh(\th+\th')+\rt{-2\mu}\ri) \cr
&
+\hf\log\lf(2\rt{-\mu}\cosh(\th-\th')+\rt{-2\mu}\ri)~.
}}
 
Let us compare \Zneuttwo\
with the semiclassical wavefunction of neutral FZZT brane
\eqn\psizerotwo{
\psi_0(x,q)\sim (-ix)^{-|q|}\exp\lf[D_0(x)-\hf\log x+\cdots\ri]~.
}
Since there are no charged ZZ-branes in this phase and
the neutral brane has no RR Ishibashi component in its boundary state,
the leading semiclassical terms ({\it i.e.} disk and annulus) 
should be $q$-independent \SeibergEI.
Therefore, the rescaled wavefunction $(-ix)^{|q|}\psi_0(x)$,
 which is independent
of $q$ at the leading order, is the natural object to compare with
the worldsheet calculation.
Using the relation $x=i\rt{-{\mu\o2}}\cosh\th$,
one can see that $Z(\si,0|\si,0)$ in \Zneuttwo\
agrees with the $-\log x$ term in \psizerotwo.

In a similar manner as in appendix A,
the annulus amplitude between the charged branes using the naive
boundary state is evaluated as
\eqn\chchintwoint{\eqalign{
Z(\si,\xi|\si',\xi')^{naive}
&=\hf\int_{-\infty}^\infty dP{P\o P^2+\vep^2}
{\cos(\pi P\si+\xi\xi'\pi P\si')\o\sinh(\rt{8}\pi P)}\cr
&={1\o2\rt{8}\vep}-\log\lf[2\cosh\lf({\th+\xi\xi'\th'\o2}
\ri)\ri] \cr
&=-\qu\log(-\mu)-\log\lf[2\cosh\lf({\th+\xi\xi'\th'\o2}
\ri)\ri]+{\rm const}~.
}}
From this expression, the amplitude between a charged FZZT-brane 
and itself is
found to be
\eqn\FtoFtwonaive{
Z(\si,\xi|\si,\xi)^{naive}=-\qu\log(-\mu)
-\log(2\cosh\th)~.
}

Now let us compare the amplitude \FtoFtwonaive\
with the semiclassical wavefunction of charged FZZT-brane \SeibergEI
\eqn\psiptwo{
\psi_+\sim \exp\lf[-{4\o3}\lf(-{\mu\o2}\ri)^{3\o2}\cosh^3\th
-\hf\log(2\cosh\th)+\hf\th+q\lf(\th+\hf\log(-\mu)\ri)\ri]~.
}
It is argued \SeibergEI\ that the annulus amplitude extracted from the
asymptotic wavefunction \psiptwo\ has a decomposition
\eqn\Zdecomp{
Z(\si|\si)=-\log(2\cosh\th)+\th=Z(\si|\si)^{naive}+\del_q D^{naive}(\si)
-\qu \del_q^2 F~,
}
where $Z(\si|\si)^{naive}$ is the naive annulus amplitude,
$\del_q D^{naive}(\si)=\th$ is the disk one-point function
of $V_R$, and $\del_q^2 F=-\log(-\mu)$
is the two-point function of $V_R$ on the sphere. 
Therefore, the matrix model result predicts that \SeibergEI
\eqn\ZDFexp{
Z(\si|\si)^{naive}=-\qu\log(-\mu)
-\log(2\cosh\th)~.
}
Clearly, our boundary state computation \FtoFtwonaive\ for the naive part 
of annulus amplitude
agrees with the matrix model result \ZDFexp\ as conjectured in \SeibergEI.

To summarize, our analysis of annulus amplitudes in section 2 and 3
precisely reproduces
the all known asymptotic behaviors of semiclassical wavefunctions studied
in \SeibergEI. This agreement between the worldsheet computation
and the matrix model result can be thought of as strong evidence
of the duality between the $(2,4)$ minimal superstring and 
the double scaled matrix model in the presence
of D-branes.

\newsec{Multi-Point Correlators of Neutral Branes}
As we saw above, only the annulus amplitude $Z(x,x)$ 
between a FZZT-brane and itself 
appears in the semiclassical expansion of the wavefunctions,
or the one-point functions of the FZZT operators.
It is clear that the natural place where the annulus amplitudes
$Z(x,x')$ with $x\not= x'$ show up
is the multi-point functions of FZZT-brane operators.
For the neutral branes, as we will see below,
it is straightforward to generalize the
one-point function $\bra B_0(x)\ket$ to
the multi-point functions $\bra\prod_i B_0(x_i)\ket$
as in the case of bosonic minimal string \refs{\MaldacenaSN,\HashimotoBF}.
On the other hand, the calculation of 
charged brane correlators turned out to be not as 
straightforward as that of neutral branes.
Therefore, in this paper we will consider only the correlators
of neutral branes and leave the matrix model
calculation of charged brane correlators
as an interesting open problem.

It is argued \SeibergEI\ that the wavefunction of neutral FZZT brane is
obtained as a double scaling limit of the determinant operator
in a complex matrix model
\eqn\detpsizero{
\psi_0(x)=\lim_{N\riya\infty}
{1\o\rt{h_N}}e^{-V(x^2)}\Big\bra\det(x^2-MM^\dag)\Big\ket=
{1\o Z}\bra B_0(x)\ket~.
}
We can naturally promote this equation as the definition of the operator
$B_0(x)$
\eqn\Bzeromatrix{
B_0(x)=\lim_{N\riya\infty}
{1\o\rt{h_N}}e^{-V(x^2)}\det(x^2-MM^\dag)~.
}
As explained in the appendix A.3 of \SeibergEI, we can apply the
determinant formula of \MorozovHH\ to evaluate the correlator of determinants
in the complex matrix model
\eqn\multipsizero{
\lf\bra\prod_i\det(x_i^2-MM^\dag)\ri\ket={\det(P_{N+j-1}(x_i^2))\o\lap(x^2)}~,
}
where $\lap(\la)=\prod_{i>j}(\la_i-\la_j)$ is the Vandermonde determinant
and $P_n(\la)$ are orthogonal polynomials.
In the double scaling limit, the shift in the index of 
the orthogonal polynomials becomes a derivative
with respect to $\mu$. Therefore, after double scaling limit
\multipsizero\ becomes
\eqn\limpsimulti{
{1\o Z}\lf\bra\prod_i B_0(x_i)\ri\ket=
{\det_{ij}\Big(\del_{\mu}^{j-1}\psi_0(x_i,\mu)\Big)
\o\lap(x^2)}~.
}
Note that 
a similar expression for the multi-brane correlator in bosonic minimal 
string theory was obtained in \refs{\MaldacenaSN,\HashimotoBF}.

From \limpsimulti, one can see that the semiclassical
expansion
of the two neutral FZZT-brane correlator has a prefactor
\eqn\neutralann{
{\del_\mu D_0(x)-\del_\mu D_0(x')\o x^2-x'^2} 
={2\o\rt{|\mu|}(\cosh\th+\cosh\th')}~.
}
Note that this expression is valid for both signs of $\mu$.
As expected, \neutralann\ exactly coincides (up to an irrelevant 
numerical factor) with
the exponential of the annulus amplitude between two neutral branes
$Z(\si,0|\si',0)$ given in \annBB\ and \Zneuttwo.
This gives another nontrivial check for the duality between 
the matrix model and the minimal superstring,
in the multi-brane sector in this case.

As we mentioned above, it would be very interesting to 
generalize the one-point functions $\bra B_\pm(x)\ket$ of charged FZZT-brane
to the multi-point functions $\bra\prod_{i}B_{\pm}(x_i)\ket$
and study their interplay with the background RR flux.

\vskip8mm
\noindent
{\bf Acknowledgments:} 
I would like to thank
Nathan Seiberg and David Shih 
for useful discussions and encouragement.
I would also like to thank Jongwon Park for discussions.

\appendix{A}{Computation of Annulus Amplitudes}
In this appendix, we will present  details of the calculation of annulus 
amplitudes \FZZTFZZT\ and \Zneuttwo\ in the text. 
See \refs{\MartinecKA,\KutasovFG} 
for similar calculations in the bosonic minimal string.

Before going into details, let us summarize some useful formulas
used in the following computation.
We need the absolute value of 
the super-Liouville wavefunctions $A_{NS}(P), A_R(P)$ given in
\ANSR:
\eqn\absANSR{
|A_{NS}(P)|^2={1\o2\sinh\pi Pb\sinh\pi Pb^{-1}},\quad
|A_{R}(P)|^2={1\o2\cosh\pi Pb\cosh\pi Pb^{-1}}~.
} 
We define the partition functions as
\eqn\deffs{\eqalign{
\eta(q)&=q^{1\o24}\prod_{n=1}^\infty(1-q^n) \cr
f_R(q)&=\rt{2}q^{1\o24}\prod_{n=1}^\infty(1+q^n) \cr
f_{{NS}}(q)&=q^{-{1\o48}}\prod_{n=1}^\infty(1+q^{n-\hf})
\cr
f_{\til{NS}}(q)&=q^{-{1\o48}}\prod_{n=1}^\infty(1-q^{n-\hf})
}}
and the following expansions for the ratio of $\eta(q)$ and $f(q)$'s
are useful:
\eqn\etafexpand{\eqalign{
&{\eta(q)\o f_{NS}(q)}=\hf\sum_{n\in{\Bbb Z}}
(-1)^{\hf n(n+1)}q^{\qu\lf(n+\hf\ri)^2}
=\sum_{k\in{\Bbb Z}}\lf[q^{(8k+1)^2\o16}-q^{(8k+3)^2\o16}\ri]\cr
&{\eta(q)\o f_R(q)}={1\o\rt{2}}\sum_{n\in{\Bbb Z}}(-1)^nq^{n^2}
\cr
&{\eta(q)\o f_{\til{NS}}(q)}=
\hf\sum_{n\in{\Bbb Z}}q^{\qu \lf(n+\hf\ri)^2}
=\sum_{k\in{\Bbb Z}}\lf[q^{(8k+1)^2\o16}+q^{(8k+3)^2\o16}\ri]~.
}}
\etafexpand\ can be easily shown by using Jacobi triple product identity.
We will also use the following summation formulas
\eqn\sumformulae{\eqalign{
\sum_{k\in{\Bbb Z}}{1\o x^2+(k+a)^2}&={\pi\o x}\cdot
{\sinh(2\pi x)\o\cosh(2\pi x)
-\cos(2\pi a)} \cr
\sum_{n\in{\Bbb Z}}{(-1)^n\o n^2+a^2}&={\pi\o a\sinh(\pi a)}~.
}}

\subsec{Neutral-Neutral Amplitude: $Z(\si,0|\si',0)$}
Let us start with the annulus amplitude 
between two neutral branes described by the boundary state 
$|\si,0,\eta=+\ket$ in \Bstateone:
\eqn\neuneuint{
Z(\si,0|\si',0)=2\int_0^\infty dP|A_{NS}(P)|^2
\cos(\pi P\si)\cos(\pi P\si')\int_0^\infty dt_c
\lf[{\eta(q)\o f_{NS}(q)}\ri]^2{f_{NS}(q)\o \eta(q)}q^{P^2\o2}~.
}
Here, the factor $[{\eta\o f_{NS}}]^2$ is the ghost contribution
and ${f_{NS}(q)\o \eta(q)}q^{P^2\o2}$
is the NS character of ${\cal N}=1$ super-Liouville theory.

Using the expansion \etafexpand\ and the summation formula 
\sumformulae, the $t_c$ integral becomes 
(recall $q=e^{-2\pi t_c}$)
\eqn\NStint{
\int_0^\infty dt_c{\eta(q)\o f_{NS}(q)}q^{P^2\o2}
={1\o\pi}\sum_{k\in{\Bbb Z}}
\lf[{1\o P^2+{(8k+1)^2\o8}}-{1\o P^2+{(8k+3)^2\o8}}\ri]
={\sinh{\pi P\o\rt{2}}\o P\cosh(\rt{2}\pi P)}~.
}
Then the amplitude \neuneuint\ becomes
\eqn\Zneusimp{\eqalign{
Z(\si,0|\si',0)&=\int_{-\infty}^\infty dP|A_{NS}(P)|^2
\cos(\pi P\si)\cos(\pi P\si'){\sinh{\pi P\o\rt{2}}\o P\cosh(\rt{2}\pi P)}\cr
&=\int_{-\infty}^\infty {dP\o P}{\cos(\pi P\si)\cos(\pi P\si')\o
\sinh(\rt{8}\pi P)}~.
}}
As discussed in \KutasovFG,
the double pole at $P=0$ can be regularized by replacing
\eqn\regularization{
{1\o P}\Riya \lim_{\vep\riya+0}{P\o P^2+\vep^2}~.
}
By closing the contour of the $P$-integral
and picking up the residues of the poles,
the integral \Zneusimp\ is evaluated as 
\eqn\Zneuform{
Z(\si,0|\si',0)
={1\o\rt{8}\vep}-\log\lf(2\cosh\th
+2\cosh\th'\ri)~.
}
We identify the divergence ${1/\vep}$ as the volume
of the Liouville direction.
Recalling 
our definition of the vertex operator ${\cal V}_P=e^{({Q\o2}+iP)\phi}$
and the closed string tachyon $T=\mu e^{b\phi}$,
$1/\vep$ is identified as
\eqn\volumeVL{
{1\o\vep}={1\o b}\log\lf({\La\o|\mu|}\ri)
}
where $\La$ is a cut-off. After removing the
divergent $\log\La$ term, we obtain
\eqn\Zneufinal{
Z(\si,0|\si',0)=-\hf\log\mu-\log\lf(2\cosh\th
+2\cosh\th'\ri)
+{\rm const}~.
}

\subsec{Charged-Charged Amplitude: $Z(\si,\xi|\si',\xi')$}
The annulus between charged FZZT branes consists of two contributions:
one from the NSNS sector and the other from
the RR sector.
From the explicit form of the boundary states \Bstateone,
one can see that
the NSNS sector exchange is the half of neutral brane amplitude
computed in the previous section
\eqn\chchNS{
Z(\si,\xi|\si',\xi')_{NS}=\hf Z(\si,0|\si',0)~.
}
On the other hand, the RR sector exchange is given by
\eqn\chchRR{
Z(\si,\xi|\si',\xi')_{R}
=-\xi\xi'\int_0^\infty dP|A_{R}(P)|^2
\cos(\pi P\si)\cos(\pi P\si')\int_0^\infty dt_c
{\eta(q)\o f_{R}(q)}q^{P^2\o2}~.
}
Here we reversed the sign of RR bra-state 
in order to impose the correct GSO projection in the open string channel.
Again, using \etafexpand\ and 
\sumformulae, the $t_c$ integral is evaluated as
\eqn\Rtint{
\int_0^\infty dt_c
{\eta(q)\o f_{R}(q)}q^{P^2\o2}={1\o2P\sinh{\pi P\o\rt{2}}}~.
}
Plugging this into \chchRR, it turns out that RR exchange is
also proportional to $Z(\si,0|\si',0)$
\eqn\chRinneut{
Z(\si,\xi|\si',\xi')_{R}=-{\xi\xi'\o2}Z(\si,0|\si',0)~.
}
Finally the total amplitude is
\eqn\totalchch{
Z(\si,\xi|\si',\xi')=Z(\si,\xi|\si',\xi')_{NS}+Z(\si,\xi|\si',\xi')_{R}
={1-\xi\xi'\o2}Z(\si,0|\si',0)~.
}

\subsec{Neutral-Charged Amplitude: $Z(\si,0|\si',\xi)$}
The annulus between a neutral FZZT-brane and a charged FZZT-brane 
is given by
\eqn\mixedone{
Z(\si,0|\si',\xi)=-\rt{2}\int_0^\infty dP|A_{NS}(P)|^2
\cos(\pi P\si)\cos(\pi P\si')\int_0^\infty dt_c
{\eta(q)\o f_{\til{NS}}(q)}q^{P^2\o2}~.
}
In the same way as before, the $t_c$ integral is found to be
\eqn\tilNStint{
\int_0^\infty dt_c
{\eta(q)\o f_{\til{NS}}(q)}q^{P^2\o2}={\tanh\rt{2}\pi P\o\rt{2}P} ~.
}
Then \mixedone\ becomes
\eqn\mixedsum{\eqalign{
Z(\si,0|\si',\xi)&=-\rt{2}\int_{-\infty}^\infty {dP\o2}
|A_{NS}(P)|^2
\cos(\pi P\si)\cos(\pi P\si'){\tanh\rt{2}\pi P\o\rt{2}P} \cr
&=-\int_{-\infty}^\infty {dP\o P}{\cos(\pi P\si)\cos(\pi P\si')\o
\sinh(\rt{8}\pi P)}\cosh{\pi P\o\rt{2}}~.
}}
Finally, regularizing the divergence using \regularization\
and closing the contour of the $P$-integral, we find
\eqn\mizedresult{\eqalign{
&Z(\si,0|\si',\xi)\cr
=&\hf\log\lf(2\rt{\mu}\cosh(\th+\th')+\rt{2\mu}\ri)
+\hf\log\lf(2\rt{\mu}\cosh(\th-\th')+\rt{2\mu}\ri) \cr
=&\hf\log\lf[\cosh\lf(\th+{\pi i\o4}\ri)+\cosh\th'\ri]
+\hf\log\lf[\cosh\lf(\th-{\pi i\o4}\ri)+\cosh\th'\ri]+\log(2\rt{\mu})~.
}}


\listrefs
\bye